\documentclass[%
12pt,
reprint,
superscriptaddress,
amsmath,amssymb,
floatfix,
]{revtex4-2}

\usepackage{graphicx}
\usepackage{rotating}
\usepackage{dcolumn}
\usepackage{bm}

\usepackage{xcolor}
\usepackage[colorlinks, allcolors=blue]{hyperref}
\usepackage{orcidlink}

\usepackage{txfonts}
\usepackage{microtype}
\usepackage{etoolbox}
\usepackage{amsmath}

\begin{document}

\title{Probing the state of hydrogen in $\delta$-AlOOH at mantle conditions with machine learning potential}

\author{Chenxing Luo\,\orcidlink{0000-0003-4116-6851}}
\affiliation{Department of Applied Physics and Applied Mathematics, Columbia University, New York, New York 10027, USA}

\author{Yang Sun\,\orcidlink{0000-0002-4344-2920}}
\email[]{yangsun@iastate.edu}
\thanks{current address: Department of Physics, Xiamen University, Xiamen, Fujian, China}
\affiliation{Department of Applied Physics and Applied Mathematics, Columbia University, New York, New York 10027, USA}
\affiliation{Department of Physics and Astronomy, Iowa State University, Ames, Iowa 50011, USA}

\author{Renata M.\ Wentzcovitch\,\orcidlink{0000-0001-5663-9426}}
\email[]{rmw2150@columbia.edu}
\affiliation{Department of Applied Physics and Applied Mathematics, Columbia University, New York, New York 10027, USA}
\affiliation{Department of Earth and Environmental Sciences, Columbia University, New York, New York 10027, USA}
\affiliation{Lamont--Doherty Earth Observatory, Columbia University, Palisades, New York 10964, USA}
\affiliation{Data Science Institute, Columbia University, New York, New York 10027, USA}
\affiliation{Center for Computational Quantum Physics, Flatiron Institute, New York, NY 10010, USA}

\date{\today}

\begin{abstract}
Hydrous and nominally anhydrous minerals (NAMs) are a fundamental class of solids of enormous significance to geophysics. They are the water carriers in the deep geological water cycle and impact structural, elastic, plastic, and thermodynamic properties and phase relations in Earth’s forming aggregates (rocks). They play a critical role in the geochemical and geophysical processes that shape the planet. Their complexity has prevented predictive calculations of their properties, but progress in materials simulations ushered by machine learning potentials is transforming this state of affairs. Here, we adopt a hybrid approach that combines deep learning potentials (DP) with the SCAN meta-GGA functional to simulate a prototypical hydrous system. We illustrate the success of this approach to simulate $\delta$-AlOOH ($\delta$), a phase capable of transporting water down to near the core-mantle boundary of the Earth ($\sim$2,900~km depth and $\sim$135~GPa) in subducting slabs. A high-throughput sampling of phase space using molecular dynamics simulations with DP-potentials sheds light on the hydrogen-bond behavior and proton diffusion at geophysical conditions. These simulations provide a pathway for a deeper understanding of these crucial components that shape Earth's internal state.
\end{abstract}

\maketitle

\section{Introduction}

H-bearing mineral phases are responsible for water circulation from the Earth's surface to the interior via subduction and convection. Up to ten oceans could be stored in the Earth's interior as hydrous phases or nominally anhydrous minerals (NAMs), where H exists as substitutional or interstitial defects \cite{ohtaniRoleWaterEarth2020, ohtaniHydrationDehydrationEarth2021}. Hydrogen bonds (H-bonds) transform under elevated pressures and temperatures in these solids. Like H-bonds in H$_2$O-ice, they symmetrize under pressure, producing ionic bonds \cite{benoitShapesProtonsHydrogen2005, zhangDeepNeuralNetwork2020} or disorder before melting \cite{umemotoOrderDisorderPhase2010, hernandezSuperionicSuperionicPhaseTransitions2016}. Superionic diffusion of protons is also expected as a precursor to dehydration reactions responsible for melt production and volcanism. In NAMs, the H-concentration changes phase relations and plastic properties, causing seismological properties irregularities in the mantle (e.g., \cite{jacobsenEffectsHydrationElastic2008, wangElasticityHydrousRingwoodite2021, wangFirstprinciplesStudyWater2022, hanPervasiveLowvelocityLayer2021}). Fragile H-bonds and hydrous defects weaken the rock's rheological properties, facilitating plastic deformation and thermal convection in the mantle, a central process in Earth's evolution (e.g., \cite{vankekenSubductionFactoryDepthdependent2011, ohtaniFateWaterTransported2018, dengElasticAnisotropyLizardite2022}). Therefore, a detailed understanding of H-bond behavior and related mineral properties is essential for understanding the Earth's interior's geochemical activity, dynamic behavior, and seismological properties.

Because the pressures and temperatures ($P,T$s) in the Earth's interior challenge the experimental determination of materials' properties, insights from \textit{ab initio} studies have been indispensable. Standard LDA \cite{perdewSelfinteractionCorrectionDensityfunctional1981} and PBE \cite{perdewGeneralizedGradientApproximation1996} functionals combined with phonon calculations and the quasiharmonic approximation (QHA) \cite{qinQhaPythonPackage2019, wuQuasiharmonicThermalElasticity2011, luoCijPythonCode2021} or with \textit{ab initio} molecular dynamics (MD) \cite{wentzcovitchFirstPrinciplesMolecular1991} have been used to address the physical properties of such phases. However, anharmonicity, small MD simulation size, inadequate DFT functionals, quantum nuclear effects, and other factors prevented the reliable description of H-bond behavior and property changes in these phases at relevant conditions.

Here, we focus on $\delta$-AlOOH ($\delta$) \cite{suzukiNewHydrousPhase2000}, a prototypical hydrous phase stable throughout the entire pressure range of the Earth's mantle (up to 135~GPa) \cite{duanPhaseStabilityThermal2018}. It has been extensively studied experimentally (e.g., \cite{sano-furukawaChangeCompressibilityDAlOOH2009, duanPhaseStabilityThermal2018, sano-furukawaDirectObservationSymmetrization2018}) and computationally (e.g., \cite{tsuchiyaFirstPrinciplesCalculation2002, tsuchiyaVibrationalPropertiesDAlOOH2008, tsuchiyaElasticPropertiesDAlOOH2009a, luoInitioInvestigationHbond2022}). Therefore, experimental uncertainties and \textit{ab initio} calculations' limitations to reproducing observations are well established. This phase has a simple high-pressure CaCl$_2$-type structure \cite{suzukiNewHydrousPhase2000, ohtaniStabilityFieldNew2001} and exhibits anomalous elastic properties under pressure \cite{tsuchiyaElasticPropertiesDAlOOH2009a}, with proton diffusion expected at high temperatures \cite{heSuperionicHydrogenEarth2018}. $\delta$, its isostructural siblings $\epsilon$-FeOOH, MgSiO$_4$H$_2$ phase H \cite{tsuchiyaFirstPrinciplesPrediction2013}, the NAM CaCl$_2$-type SiO$_2$, along with their multiple solid solutions, are the main H-bearing phases in the deep mantle \cite{tsuchiyaInitioStudyLower2020}. Although $\delta$-AlOOH is not the most dominant hydrous phase in the lower mantle or core, the structural simplicity, the abundance of careful measurements and calculations of $\delta$'s properties, and its prototypical nature make it a most suitable model phase for testing the performance of \textit{ab initio}-based machine learning (ML) methods in addressing these systems at high $P,T$s.

According to measurements, H-bonds in $\delta$ symmetrize and form H-centered (HC) bonds at $\sim$18~GPa \cite{sano-furukawaChangeCompressibilityDAlOOH2009, sano-furukawaDirectObservationSymmetrization2018}.
Quasiharmonic approximation (QHA) based methods (e.g., \cite{qinQhaPythonPackage2019, wuQuasiharmonicThermalElasticity2011, luoCijPythonCode2021}), while widely used to describe finite-temperature properties of solids, rely on stable phonon modes.
While these methods describe effects caused by the anomalous pressure dependence of OH-stretching modes observed in the infrared (IR) spectroscopy below 12~GPa \cite{kagiInfraredAbsorptionSpectra2010, luoInitioInvestigationHbond2022}, they cannot describe $\delta$'s properties in the pressure range of H-bond symmetrization, a transition accompanied by strong anharmonicity and an order-disorder precursor below $\sim$40~GPa in static calculations \cite{tsuchiyaVibrationalPropertiesDAlOOH2008, luoInitioInvestigationHbond2022}. Besides, previous \textit{ab initio} QHA studies have used the LDA and PBE/GGA functionals, making comparing \textit{ab initio} predictions and experimental measurements at the same pressures challenging. The relatively low experimental H-bond symmetrization pressure of $\sim$8--18~GPa has not been correctly reproduced in these DFT calculations because of the process's complexity and these functional's inadequate description of H-bonds, giving a transition pressure in the 30--40~GPa pressure range \cite{luoInitioInvestigationHbond2022, sano-furukawaDirectObservationSymmetrization2018}.

At high temperatures, superionic proton diffusion is anticipated in H-bearing systems (e.g., antigorite, diopside with H defects), usually as a precursor to dehydration. Diffusion, critical to understanding dehydration dynamics,
has also been found in other H-bearing systems like antigorite \cite{sawaiDehydrationKineticsAntigorite2013}, diopside with H defects \cite{ingrinDiffusionHydrogenDiopside1995}, and FeOOH \cite{houSuperionicIronOxide2021} and has been extensively investigated in ice before melting (e.g., \cite{cavazzoniSuperionicMetallicStates1999, katohProtonicDiffusionHighPressure2002, goldmanBondingSuperionicPhase2005, wilsonSuperionicSuperionicPhase2013, sunPhaseDiagramHighpressure2015, millotNanosecondXrayDiffraction2019}). A recent Born-Oppenheimer molecular dynamics (BOMD) study using the PBE functional reported proton diffusion in $\delta$ at 2,700--3,000~K \cite{heSuperionicHydrogenEarth2018}, a temperature range higher than the 1,600--2,500~K dissociation ($2 \, \mathrm{AlOOH} \rightarrow \mathrm{H}_2\mathrm{O} + \mathrm{Al}_2\mathrm{O}_3$) temperature measured in the 20--140~GPa range \cite{ohtaniStabilityFieldNew2001, duanPhaseStabilityThermal2018, pietDehydrationDAlOOHEarth2020}. The PBE functional and the small MD simulation cell sizes with only a few hundred atoms are likely responsible for this discrepancy. In summary, accurately reproducing $P,T$ stability fields of hydrous phases and their properties is fundamental for mapping out phase relations and dehydration sequences along a specific geotherm (e.g., \cite{vankekenSubductionFactoryDepthdependent2011}). These results provide the basis for interpreting irregular seismic properties caused by the non-homogeneous distribution of these H-bearing solids and melts in the mantle and estimating the water content in Earth's interior (e.g., \cite{ohtaniFateWaterTransported2018, ohtaniRoleWaterEarth2020, ohtaniHydrationDehydrationEarth2021}).

This challenge can be overcome by adopting ML-based potentials (e.g., \cite{behlerGeneralizedNeuralNetworkRepresentation2007, bartokGaussianApproximationPotentials2010, thompsonSpectralNeighborAnalysis2015, shapeevMomentTensorPotentials2016, artrithEfficientAccurateMachinelearning2017, zhangEndtoendSymmetryPreserving2018}) in MD simulations with DFT predictive power but significantly reduced computational cost \cite{zuoPerformanceCostAssessment2020}. This approach only invokes DFT while training the potential to reproduce interatomic forces and energies using ML-based descriptors. The ML-based MD simulations are then performed at a cost and scaling close to empirical force-field simulations.
Active learning schemes (e.g., \cite{smithLessMoreSampling2018, zhangActiveLearningUniformly2019}) help create compact reference datasets by iteratively discovering and actively adding ``missing'' necessary atomic configurations, reducing the DFT computation cost in the potential training process. 
In particular, \textsc{Deep Potential} (DP) is a neural-network (NN) type ML potential \cite{behlerGeneralizedNeuralNetworkRepresentation2007} with a neural network's flexible fitting capability, allowing NN-potentials to represent chemical systems of varied nature.
Recent benchmarks have shown that DP forces and energies in solids and liquids structures trained on a few thousand reference configurations can be highly accurate \cite{zuoPerformanceCostAssessment2020, zhangDeepMachineLearning2022, wanThermoelasticPropertiesBridgmanite2024}. The adoption of ML-based potentials allows the use of the strongly constrained and appropriately normed (SCAN) meta-generalized gradient approximation (meta-GGA) functional's more accurate description of the H-bonded systems \cite{sunAccurateFirstprinciplesStructures2016} regardless of being computationally more expensive. This method successfully calculated the water/ice phase diagram \cite{zhangPhaseDiagramDeep2021} and proton diffusion in liquid water \cite{zhangModelingLiquidWater2021}.

Combining deep learning potentials with the SCAN functional is a promising path to simulate H-bearing systems at extreme geophysical conditions accurately. Here, we apply this hybrid approach to $\delta$, a prototypical H-bearing system, to understand how well it overcomes the limitations presented in conventional purely DFT studies. This method enables us to perform long and large simulations that densely cover a wide $(P,T)$ or $(V,T)$ range with changing H-bond or proton diffusion behavior. We benchmark the potential against SCAN calculations and experimental measurements to assess the accuracy level achieved. We address $\delta$'s compression curve in its high-$P$ H-bond symmetrization regime and its high-$T$ proton diffusion regime, which were not accurately described in previous QHA- or BOMD-based studies.

\section{Result}

\subsection{NN-potential benchmark}
\label{sec:potential}

\begin{figure}[htbp]
    \centering
    \includegraphics[width=.45\textwidth]{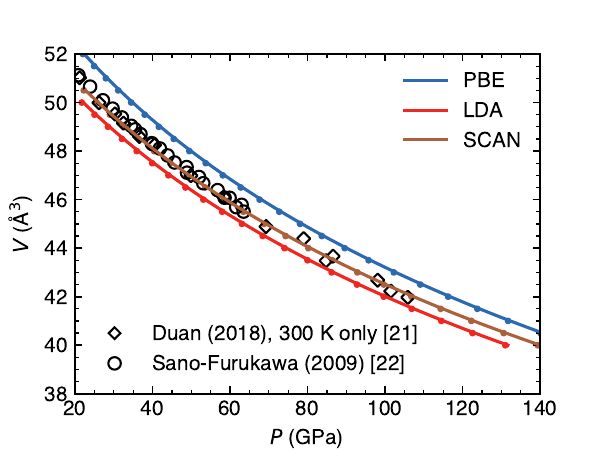}
    \caption{LDA, PBE, and SCAN static compression curves compared to 300~K measurements \cite{duanPhaseStabilityThermal2018, sano-furukawaChangeCompressibilityDAlOOH2009}.}
    \label{fig:1}
\end{figure}

We employ the SCAN functional \cite{sunAccurateFirstprinciplesStructures2016} to describe $\delta$'s potential energy (Born-Oppenheimer) surface and interatomic forces.
Compared to LDA's underestimation and PBE's overestimation of the pressure, SCAN's static compression curve is the closest to the 300~K measured one (see Fig.~\ref{fig:1}).
The slight underestimation of the SCAN volume is eliminated when thermal effects at 300~K are included, as shown in the subsequent discussion.
Quantum nuclear effects disregarded in the present study might spoil this good agreement \cite{umemotoNatureVolumeIsotope2015}, and this effect must be further inspected. We are interested in the temperature range of typical subducting tectonic plates (slabs), i.e., along ``slab geotherms'' \cite{eberleNumericalStudyInteraction2002, ohtaniHydrousMineralsStorage2015}, where this phase is stable in the mantle, i.e., $\sim\!\mbox{1,000~K} < T < \mbox{3,000~K}$ \cite{litasovPhaseRelationsHydrous2005, duanPhaseStabilityThermal2018}. At these temperatures, classical MD should adequately describe the ionic dynamics in $\delta$.

\begin{figure*}[htbp]
    \centering
    \includegraphics[width=.6\textwidth]{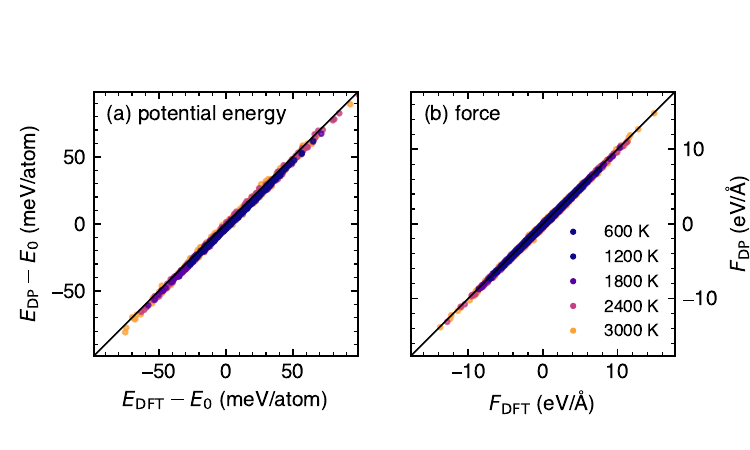}
    \caption{Comparison between SCAN-DP and SCAN-DFT predictions of (a) 1,600~potential energies and (b) 10,240~atomic forces randomly sampled from 30,128 configurations derived from 20~SCAN-BOMD $NVT$ trajectories at 5~temperatures ($T =$ 600, 1,200, 1,800, and 3,000~K) and 4~volumes ($V =$ 40, 44, 48, 54~\AA$^3$). Symbol colors denote temperature.}
    \label{fig:2}
\end{figure*}

Our SCAN-DP interatomic potential trained on just a few thousand reference configurations can reproduce SCAN-DFT's forces and energies for configurations sampled from SCAN-BOMD simulations with 128 atoms, as illustrated in Fig.~\ref{fig:2}.
The root-mean-square error (RMSE) \cite{behlerFourGenerationsHighDimensional2021} of the SCAN-DP predictions for these tests at various $(V,T)$ states are listed in Table~SI. Configurations generated at higher temperatures produce larger RMSEs in general. At 3,000~K, the RMSEs for potential energy and force predictions are $\sim$2~meV/atom, and $\sim$0.12~eV/Å, respectively. Such accuracy is similar to previous benchmarks (e.g., \cite{zuoPerformanceCostAssessment2020}) in similar DP studies.

\begin{figure}[htbp]
    \centering
    \includegraphics[width=.45\textwidth]{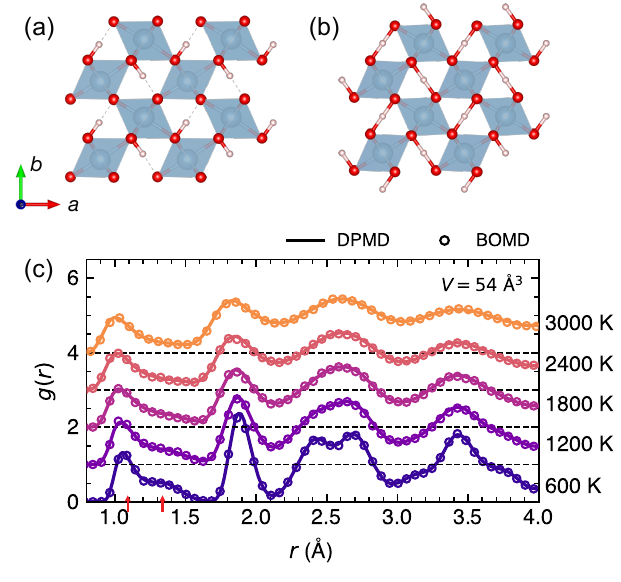}
    \caption{(a, b) The crystal structures of $\delta$-AlOOH with (a) asymmetric and (b) symmetric H-bonds. (c) Comparison between SCAN-BOMD and SCAN-DPMD pair-distribution function, $g(r)$, at various $(T,V)$'s. The conventional-cell volume 54~Å$^3$ corresponds to 9.42, 13.67, 17.90, 22.73, and 28.34~GPa at 600--3,000~K. Red arrows show the H-bond and OH ionic bond lengths in the first $g(r)$ peak.}
    \label{fig:3}
\end{figure}

SCAN-DPMD also reproduces SCAN-BOMD's structure and bonding properties at high-$P,T$, as illustrated by the comparison of SCAN-BOMD's and SCAN-DPMD's radial distribution functions, $g(r)$, in Fig.~\ref{fig:3}. The results are for 54~Å$^3$ volume, corresponding to 9.5--28.3~GPa at 600--3,000~K obtained in 128-atom simulations. Plots detailing the contribution from each type of atomic pair at other $P,T$'s are shown in Fig.~S1 \footnote{See Supplemental Material online for Tables SI and SII, Figs.~S1--S8.}.

The $g(r)$ at $V =$ 54~Å$^3$ shown in Fig.~\ref{fig:3} is particularly significant because this volume approximately corresponds to the critical volume of the experimentally observed H-bond disorder transition at 300~K \cite{sano-furukawaChangeCompressibilityDAlOOH2009}.
At 600~K, the first peak at $\sim$1.1~Å corresponds to the ionic OH bond length. This peak is asymmetric at this volume and at all sampled temperatures due to asymmetric proton motion. At 600~K, $g(r)$ shows a near double-peak distribution of OH ionic bonds and H-bonds; the broad shoulder centered at $\sim$1.4~Å corresponds to the distribution of H-bond lengths. The increased overlap of the two peaks with increasing temperature suggests the onset of a disordered state, a precursor to H-bond symmetrization \cite{sano-furukawaDirectObservationSymmetrization2018}. A similar OH bond-length distribution was observed in 300~K classical BOMD simulations with a quantum thermostat \cite{bronsteinThermalNuclearQuantum2017}.
Some features on $g(r)$ disappear at elevated temperatures. For example, the 2.4~Å peak, a combination of Al-O and O-H contributions, and the 2.7~Å peak associated with O-O bonds (see Fig.~S1) are split at 600~K but merge above 1,200~K. This phenomenon results from increased atomic vibration amplitudes, proton dynamic disorder, and diffusion onset at elevated temperatures. We will elaborate further on the diffusion process in Sec.~\ref{sec:simulation}.
Under pressure, the first peak at $\sim$1.1~Å and the shoulder at 1.4~Å evolve into a single symmetric peak due to H-bond disorder and finally symmetrization \cite{sano-furukawaDirectObservationSymmetrization2018, luoInitioInvestigationHbond2022} (see Fig.~S1).

\begin{figure*}[htbp]
    \centering
    \includegraphics[width=.70\textwidth]{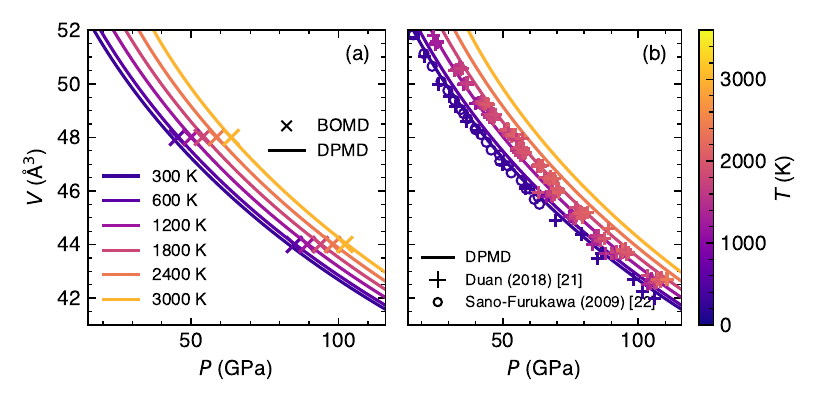}
    \caption{$\delta$'s SCAN-DPMD compression curves at various temperatures compared to (a) SCAN-BOMD's $PVT$ data and (b) measurements \cite{duanPhaseStabilityThermal2018, sano-furukawaChangeCompressibilityDAlOOH2009}. Volumes correspond to the conventional unit cell containing two~f.u.\ of $\delta$-AlOOH.}
    \label{fig:4}
\end{figure*}

SCAN-DPMD with \textit{ab initio}-level accuracy and improved efficiency will help predict the thermoelastic properties of hydrous solids, a property of first-order importance in geophysics \cite{wanThermoelasticPropertiesBridgmanite2024, wentzcovitchFirstPrinciplesQuasiharmonic2010, luoCijPythonCode2021}. Here, we compare its predictions for $\delta$'s high-temperature compressive behavior with \textit{ab initio} SCAN-BOMD predictions and measurements in the 300--3,000~K and 20--115~GPa range. Fig.~\ref{fig:4}(a) shows excellent agreement between SCAN-DPMD's high-temperature compression curves fitted to third-order Birch-Murnaghan equations of state (EoS) and the SCAN-BOMD's predicted $P,V$ data points.
Comparison with measurements \cite{duanPhaseStabilityThermal2018, sano-furukawaChangeCompressibilityDAlOOH2009} in Fig.~\ref{fig:4}(b) shows a promising 3~GPa maximum error in pressure prediction at the same $V,T$s. The next section will present a more detailed analysis of the 300~K compression curve at lower pressure and compare it with measurements.

These good agreements between the SCAN-DFT and SCAN-DP predictions of forces, potential energies, $g(r)$, and high-$P,T$ EoSs suggest that the deep-learning potentials have reached satisfactory accuracy and are predictive at least in the 20--120~GPa pressure range and up to 3,000~K.
   
\subsection{The 300~K compression curve and dynamic stabilization of the HC phase}

\begin{figure*}
    \centering
    \includegraphics[width=0.90\textwidth]{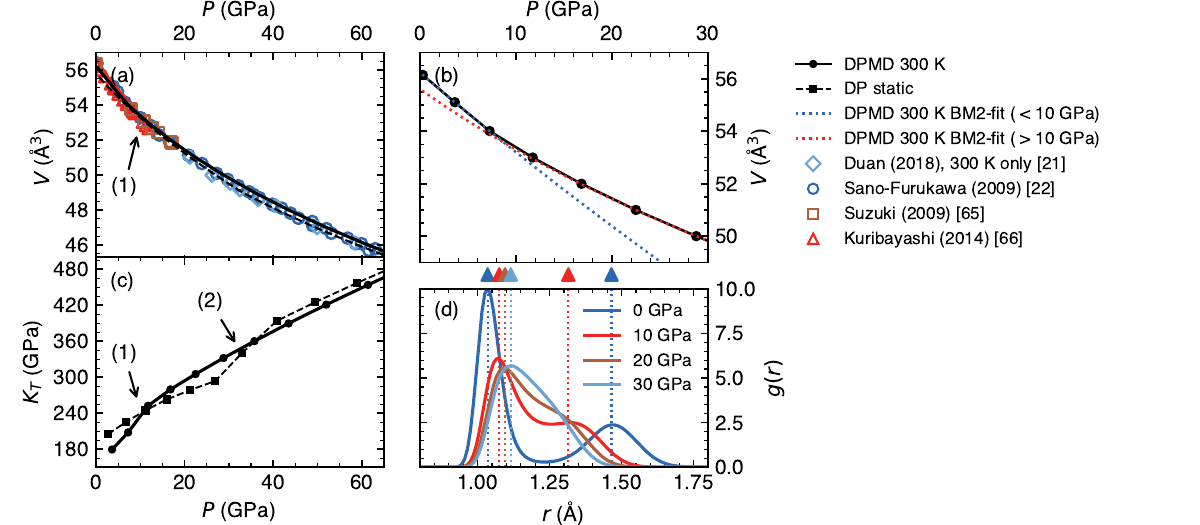}
    \caption{(a) 300~K and static compression curve results are compared to 300~K measurements \cite{sano-furukawaChangeCompressibilityDAlOOH2009, suzukiCompressibilityHighpressurePolymorph2009, kuribayashiObservationPressureinducedPhase2014, duanPhaseStabilityThermal2018}. (b) Static and 300~K SCAN-DPMD results and 300~K Birch-Murnaghan EoS fitted for 10--64 GPa data. (c) 300~K and static bulk modulus ($K = -V \, \partial P / \partial V $). Arrows (1) and (2) highlight the change in compressibility at (1) 300~K at $\sim$10~GPa and (2) static conditions at $\sim$35~GPa. (d) The evolution of $r(\mathrm{OH})$ bond length distributions, $g_\mathrm{OH}(r)$, at 0 (blue), 10 (red), and 20~GPa (brown). Color dashed lines and filled triangles indicate the $r$ of the peaks.}
    \label{fig:5}
\end{figure*}

\begin{figure}
    \centering
    \includegraphics[width=0.48\textwidth]{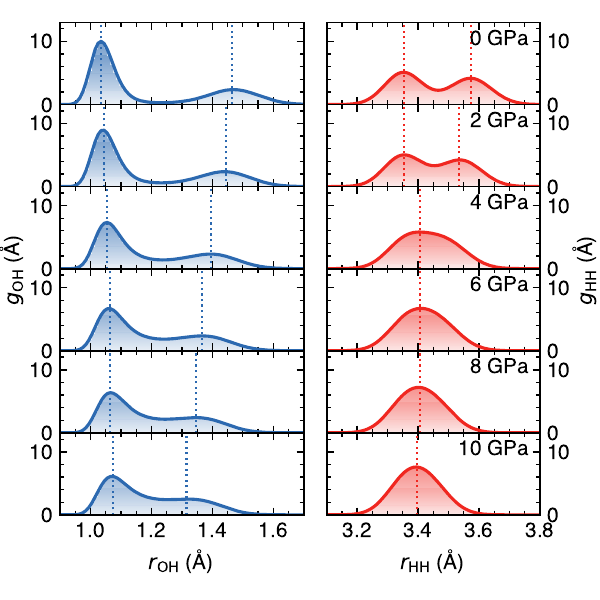}
    \caption{Pair correlation function $g_\mathrm{OH}(r)$ and $g_\mathrm{HH}(r)$ in 0--10~GPa at 300~K. Color dashed lines indicate peak extremes.}
    \label{fig:6}
\end{figure}

The 300~K compression curve is one of $\delta$'s best-determined properties.
Experimentally, $\delta$ exhibits an anomalous change in compressive behavior at $\sim$10~GPa, which is usually attributed to a change in H-bond structure, most often with H-bond symmetrization \cite{kangHydrogenBondSymmetrizationDAlOOH2017, pillaiFirstPrinciplesStudy2018, mashinoSoundVelocitiesDAlOOH2016, tsuchiyaFirstPrinciplesCalculation2002}, but also with H-bond disorder \cite{sano-furukawaChangeCompressibilityDAlOOH2009, sano-furukawaDirectObservationSymmetrization2018, simonovaStructuralStudyDAlOOH2020}.
However, it has been impossible to reproduce this compressive behavior using QHA-based DFT calculations. This is because (a)~static \textit{ab initio} calculations without multiconfiguration analysis do not account for H-bond disorder; (b)~the HC phase atomic configuration is dynamically stable above $\sim$35~GPa only in static GGA/PBE calculations \cite{tsuchiyaVibrationalPropertiesDAlOOH2008}. Strong anharmonicity produces unstable phonon modes at lower pressures, hindering the application of the QHA for high-temperature calculations \cite{luoInitioInvestigationHbond2022, tsuchiyaVibrationalPropertiesDAlOOH2008}.
Using SCAN-DPMD, we optimize and equilibrate the cell shape at several pressures to obtain the static and the 300~K compression curves shown in Figs.~\ref{fig:5}(a,b).
With these, we can analyze the effect of atomic motion on the H-bond state, its effect on the volume, and compare the latter with measurements \cite{suzukiCompressibilityHighpressurePolymorph2009, sano-furukawaChangeCompressibilityDAlOOH2009, duanPhaseStabilityThermal2018}.
As shown in Fig.~\ref{fig:5}(b), extrapolations of a Birch-Murnaghan EoS fit to results below and above $\sim$10~GPa produce divergent curves (dotted blue and red dotted lines), indicating a change in the OH-bond state at $\sim$10~GPa.

The 300~K isothermal bulk modulus, $K_T = V \, (\partial P / \partial V)_T$, shown in Fig.~\ref{fig:5}(c), amplifies the subtle change in compressibility across this critical point.
Across this state change in the 0--20~GPa range, the OH bond-length distribution, $g_\mathrm{OH}(r)$, evolves from a double peak to a single modal distribution, though not symmetric (Fig.~\ref{fig:5}(d)).
In this pressure range, the H-bond compresses quickly while the ionic OH bond stretches, a well-known behavior of these bonds (e.g., \cite{benoitQuantumEffectsPhase1998, umemotoNatureVolumeIsotope2015}).
$\delta$'s compressive behavior obtained from SCAN-DP calculation confirms previous PBE/GGA results \cite{tsuchiyaElasticPropertiesDAlOOH2009a} that this change in compressive behavior occurs in the 30--40~GPa range in static calculations (dashed line in Fig.~\ref{fig:5}(c)).

SCAN-DPMD results at 300~K account for the thermal expansion and reproduce well the 300~K experimental compression curve \cite{suzukiCompressibilityHighpressurePolymorph2009, sano-furukawaChangeCompressibilityDAlOOH2009, duanPhaseStabilityThermal2018}: in the 0--15~GPa range, the difference between results and measurements is less than the differences between measurements; at higher pressures, the predicted SCAN-DPMD volume is slightly overestimated, with an error smaller than 0.25~\AA$^3$/f.u.
Although predicted \cite{tsuchiyaFirstPrinciplesCalculation2002}, it is striking to see that the inclusion of \textit{classical} ionic motion lowers the transition pressure by $\sim$27~GPa, giving a transition pressure that agrees quite well with the experimentally measured one at $\sim$9~GPa (e.g., \cite{sano-furukawaChangeCompressibilityDAlOOH2009, mashinoSoundVelocitiesDAlOOH2016}). This 300~K result is surprising and puzzling. At this low temperature, quantum proton motion, particularly tunneling \cite{benoitTunnellingZeropointMotion1998}, should impact the H-bond behavior (e.g., \cite{wangQuantumStructuralFluxion2023}). 

Above 10~GPa, $V$ and $K_T$ display a smooth monotonic dependence on pressure.
Table~\ref{tab:eos} summarizes the Birch-Murnaghan EoS parameters resulting from the fitting of measured and calculated 300~K $P,V$ data above 10~GPa. 
Fitting the SCAN-DPMD results within the 10--64~GPa pressure range with $K'_0 = 4$ gives $V_0 = 55.5$~Å$^3$, $K_0 = 225\pm2$~GPa (red dotted line in Fig.~\ref{fig:5}(b)). The predicted $V_0$ is in excellent agreement with the measured $V_0$ \cite{sano-furukawaChangeCompressibilityDAlOOH2009} fit to the same Birch-Murnaghan EoS, 55.47~Å$^3$. The predicted bulk modulus differs by $\sim$3\% from the experimental one, 219~GPa \cite{sano-furukawaChangeCompressibilityDAlOOH2009}, an impressive agreement for a result extrapolated to 0~GPa.

Our SCAN-DPMD results are significantly more accurate than previous PBE-BOMD results, which reported a $V_0 = 58.5$~Å$^3$ and $K_0 = 183.4$~GPa using the Vinet EoS fit to results above 10~GPa \cite{bronsteinThermalNuclearQuantum2017}. This improvement can be attributed to (a)~SCAN's more accurate description of the $P$-$V$ relation, (b)~the denser sampling of $(P,V)$ states over a broader pressure range, and (c)~larger, longer, and better-converged simulation runs enabled by DPMD's performance leap compared to BOMD. Nevertheless, this level of agreement between \textit{classical} MD results and measurements at 300 K is unexpected.

This good agreement between SCAN-DPMD results and experimental data above 10~GPa suggests that our simulations describe well the H-bond state above this pressure.  As indicated in Fig.~\ref{fig:5}(d), not even at 20~GPa, the first two peaks in the $g_\mathrm{OH}(r)$ pair distribution function have merged, indicating that H-bond symmetrization has not yet been achieved in the simulations. Therefore, the change in compressive behavior at $\sim$10~GPa cannot be attributed to H-bond symmetrization in this classical simulation. Fig.~\ref{fig:6} shows in detail the evolution of $g_\mathrm{OH}(r)$ and $g_\mathrm{HH}(r)$ pair distribution functions from 0 to 10~GPa. Both show two clear peaks at 0~GPa and tend to merge with increasing pressure. However, only the $g_\mathrm{HH}(r)$ peaks are fully merged at 10~GPa. $g_\mathrm{OH}(r)$ still displays two superposing broad peaks. This indicates that H-bonds are not symmetric in these classical simulations at 10~GPa (or 20~GPa). The change in the H-bonds' state at 10~GPa seems related to a change in H-ordering, likely into a disordered state. Our previous multi-configuration PBE-QHA study \cite{luoInitioInvestigationHbond2022} suggested several short- to medium-range ordered H-bonded configurations remain dynamically stable beyond the pressure where the compressive behavior changes ($\sim$10~GPa in experiments, $\sim$12~GPa in PBE-QHA calculations). This result supported the notion of a more disordered H-bond state. The current results showing an asymmetric first peak in $g_\mathrm{OH}(r)$ and the fully merged broad peak in $g_\mathrm{HH}(r)$, and the previously identified multi-configuration equilibrium state \cite{luoInitioInvestigationHbond2022}  suggest that the change in compressive behavior at $\sim$10~GPa is associated with the emergence of a disordered state in the simulations. Static or dynamic, disorder involves proton hopping and is a precursor to proton diffusion at higher temperatures. This 300~K behavior might change significantly if the proton dynamics is addressed quantum mechanically using path integrals \cite{wangQuantumStructuralFluxion2023}.

\begin{table*}[htb]
    \centering
    \caption{Comparision of 300~K EoS parameters for HC-$\delta$.}
    \label{tab:eos}
    \begin{tabular}{rllll}
    \toprule
         & $V_0$ (\AA$^3$) & $K_0$ (GPa) & $K'_0$ & Notes \\
    \hline
    Sano-Furukawa~\textit{et al.}~\cite{sano-furukawaChangeCompressibilityDAlOOH2009} & 55.47 & 219 & 4 (fixed) & 10--63.5~GPa \\
    Duan~\textit{et al.}~\cite{duanPhaseStabilityThermal2018} & 55.3 & 223 & 4 (fixed) & Mie-Gr\"uneisen, up to 142~GPa \\
    Mashino~\textit{et al.}~\cite{mashinoSoundVelocitiesDAlOOH2016} & 56.374 & 190 & 3.7  & \\
    Simonova~\textit{et al.}~\cite{simonovaStructuralStudyDAlOOH2020} & 55.56 & 216 & 4 (fixed)  & \\
    \hline
    Bronstein~\textit{et al.}~\cite{bronsteinThermalNuclearQuantum2017} & 58.5 & 183.4 & & PBE-BOMD \\
    Kang~\textit{et al.}~\cite{kangHydrogenBondSymmetrizationDAlOOH2017} & 57.6 & 196 & 4.0 & PBE-BOMD, 20--35~GPa \\
    This study & 55.5 & $225 \pm 2$ & 4 (fixed) & SCAN-DPMD, 10--64~GPa \\
    \toprule
    \end{tabular}
\end{table*}

\subsection{Proton diffusion at high temperatures}
\label{sec:simulation}

The DP potential allows us to systematically perform large-scale DPMD simulations to investigate proton diffusion in $\delta$ over a wide $P,T$ range. We perform 96~$NVT$ DPMD simulations covering the pressure range of 35--140~GPa and temperature range of 1,500--3,000~K (see Fig.~S2). They contain 8,192~atoms and run for 2~ns.
We do not observe diffusion, structural change, or melting of the Al and O sub-lattices in this $P,T$ range.

\begin{figure}[htbp]
    \centering
    \includegraphics[width=.48\textwidth]{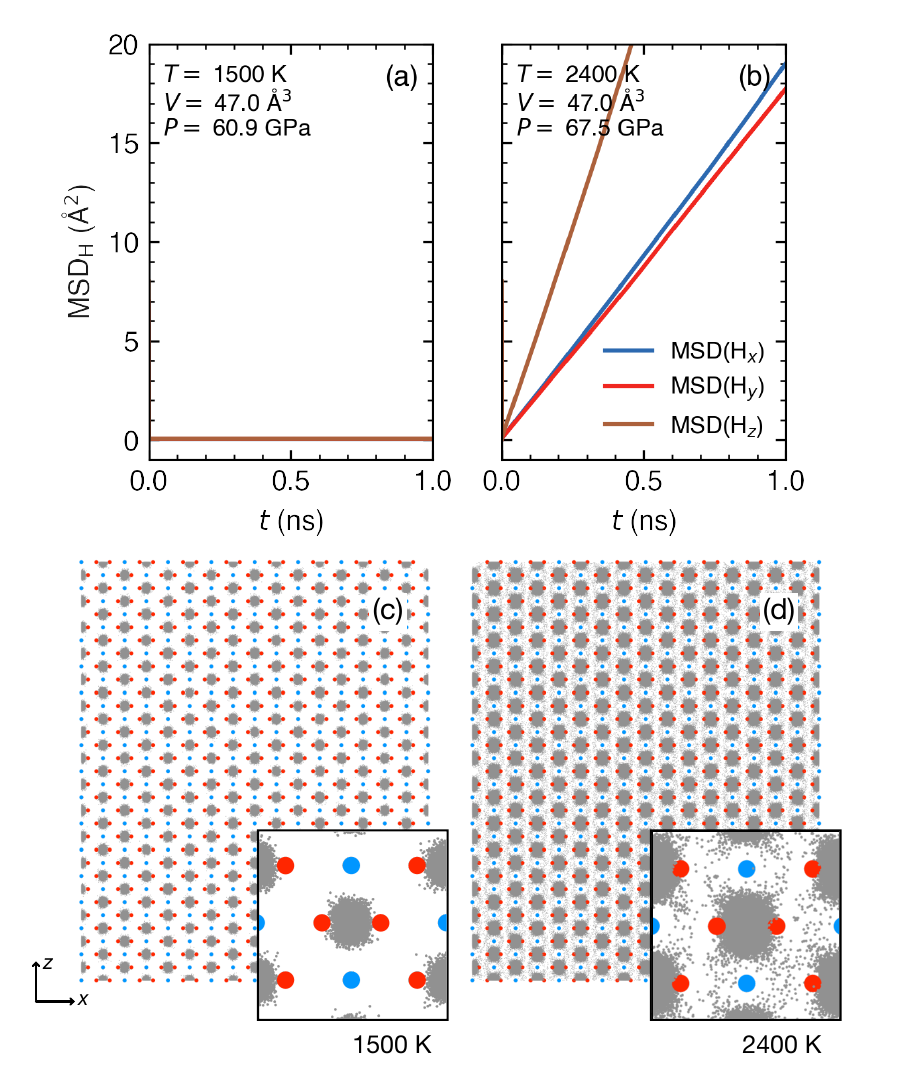}
    \caption{Proton diffusion at high temperatures for $V = 47$~\AA$^3$ conventional cell which corresponds to 60.9 and 67.5~GPa at 1,500~K and 2,400~K. (a) Proton mean-square displacement (MSD) at 1,500~K and (b) 2,400~K. (c) Proton trajectories at 1,500~K and (d) 2,400~K; blue, red, and gray dots denote Al, O, and H ions.}
    \label{fig:7}
\end{figure}

Proton diffusion is quantitatively characterized by protons' mean-square displacement (MSD). Protons' MSD over a simulation run time $t$, $L_\mathrm{H}^2(t)$, is defined by \cite{huangInitioMolecularDynamics2019, tuckermanStatisticalMechanicsTheory2010}
\begin{equation}
L_\mathrm{H}^2 (t) = \frac{1}{N_\mathrm{H}} \bigg\langle \sum_{i=1}^{N_\mathrm{H}} \big\lvert R_{\mathrm{H},i}(t+\tau) - R_{\mathrm{H},i}(\tau) \big\rvert^2\bigg\rangle,
\end{equation}
where $N_\mathrm{H}$ is the total number of protons, $R_{\mathrm{H},i}(\tau)$ denotes the position of $i$-th proton at moment $\tau$, and ``$\langle \,\cdot\, \rangle$'' denotes the ensemble average over the start time $\tau$. A linear (non-linear) dependence of the MSD on the simulation run time, $t$, reflects a continuous (irregular) diffusive behavior. The self-diffusion coefficient, $D_\mathrm{H}$, or the diffusivity, is related to the slope of the MSD ($L^2$) vs.\ $t$ line,
\begin{equation}
D_\mathrm{H} = \lim_{t\to\infty} \frac{L_\mathrm{H}^2}{6t}\,.
\end{equation}

Figs.~\ref{fig:7}(a,b) show the protons' MSD over a time span of 1~ns for $V = 47$~Å$^3$ and $T =$~1,500~K (60.9~GPa) and 2,400~K (67.5~GPa), respectively. MSD at several other $V,T$ conditions are available in Fig.~S3. Increasing the temperature significantly increases protons' diffusivity (see Fig.~S3). Diffusion is not observed at 1,500~K. It starts between 1,800~K and 2,100~K and becomes steady at 2,400~K. Figs.~\ref{fig:7}(c,d) show the corresponding proton trajectories. Fig.~\ref{fig:7}(b), shows that the protons' MSD along the $c$ direction is approximately twice as large as those along the $a$ and $b$ axes, indicating they move more freely through the interstitial channels along the $c$ direction (see Figs.~\ref{fig:3}(a,b)). A similar phenomenon has also been reported in the NAM phase hydrous Al-bearing stishovite, which has a similar CaCl$_2$-type Si-O framework \cite{umemotoPoststishoviteTransitionHydrous2016}.

\begin{figure*}[htbp]
    \includegraphics[width=.9\textwidth]{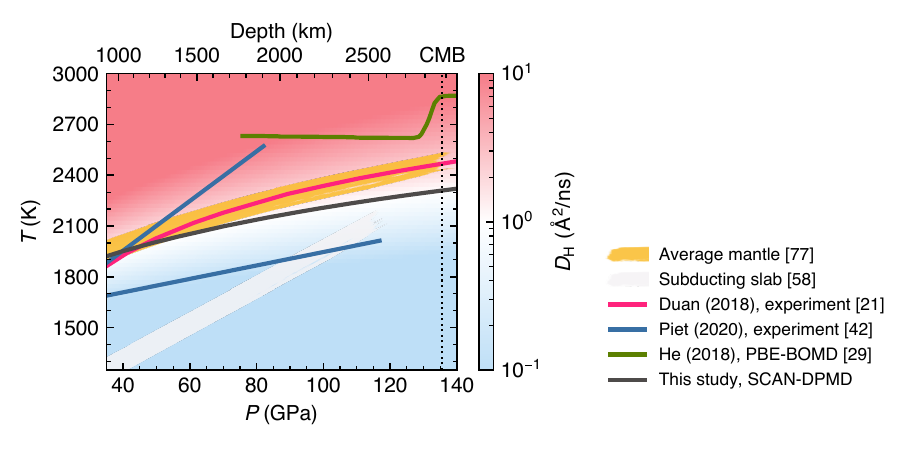}
    \caption{
    \raggedright
    The diffusion phase diagram of $\delta$.
    The background color represents the diffusivity $D_\mathrm{H}$. The solid black curve indicates the $D_\mathrm{H} = 1$~Å$^2$/ns boundary for steady proton diffusion characterized by a linear dependence of MSD vs.\ time. The diffusion boundary from Ref.~\cite{heSuperionicHydrogenEarth2018} and dissociation boundaries from Refs.~\cite{duanPhaseStabilityThermal2018, pietDehydrationDAlOOHEarth2020} are also shown.
    Adiabatic mantle geotherm \cite{brownThermodynamicParametersEarth1981} and slab geotherm \cite{eberleNumericalStudyInteraction2002}.
    }
    \label{fig:8}
\end{figure*}

At 1,800--3,000~K, the relationship between $\log D_\mathrm{H}$ and $1/T$ is linear at all volumes and can be fitted using the Arrhenius equation (see Fig.~S4),
\begin{equation}
D_\mathrm{H}(T) = D_0 \, \exp\bigg(\frac{-E_\mathrm{a}}{k_\mathrm{B}T}\bigg),    
\end{equation}
where $k_\mathrm{B}$ is the Boltzmann constant and $E_\mathrm{a}$ denotes the activation energy. This behavior indicates that proton diffusion in $\delta$ is a common thermally activated process. The volume dependence of the activation energy, $E_\mathrm{a}$, is nearly linear (Fig.~S5). Using these relationships, we interpolate $D_\mathrm{H}$ vs.\ $V$ and $T$ using the finite temperature EoS displayed in Fig.~\ref{fig:4}. The magnitude of protons' diffusivity is shown as the background color in Fig.~\ref{fig:8}.
For diffusivity corresponding to MSD $>1~\mbox{\AA}^2/\mbox{ns}$, (e.g., Fig.~7), diffusion is steady, stable, and characterized by a linear dependence of the MSD vs.\ time (reddish background area); for diffusivity of $<0.1~\mbox{\AA}^2/\mbox{ns}$ (bluish background area), diffusion can be intermittent with non-linear dependence of the MSD vs.\ time and the system is considered to be solid \cite{devillaDoubleSuperionicityIcy2023}.
Depending on the pressure, for $\mbox{1,800}~\mathrm{K} < T < \mbox{2,100}~\mathrm{K}$, $D_\mathrm{H}$ approaches 1~Å$^2$/ns and starts deviating from the high-$T$ Arrhenius behavior (see Fig.~S4). This region of the diffusivity diagram is transiting from a blueish to a whitish background. At these $P,T$ conditions, one can recognize in Fig.~S3 a change in the MSD's behavior from non-linear to linear. We designate this regime as the onset of the fully superionic behavior in the $P,T$ phase space and represent this $D_\mathrm{H} = 1$~Å$^2$/ns diffusion boundary in Fig.~\ref{fig:8} with a solid black curve.

The experimental dehydration boundary reported in Ref.~\cite{duanPhaseStabilityThermal2018} resembles our diffusion boundary.
Similar measurements in Ref.~\cite{pietDehydrationDAlOOHEarth2020} pinpointed two $\delta$ dehydration temperatures in two distinct pressure ranges: 1,850--1,900~K at 28--68~GPa and $\sim$1,959~K at 100--110~GPa. Two divergent boundaries were reported in the mid-lower mantle (68--100~GPa) due to uncertainties related to unidentified X-ray diffraction peaks and uncertain experimental conditions (Fig.~\ref{fig:8}).
Their lower boundary directly connects the low-pressure and high-pressure boundaries and runs parallel to our boundary with a 200~K downward shift.
It has been argued \cite{pietDehydrationDAlOOHEarth2020} that the disagreement between these studies could be due to uncertainties in granularity or the preferred orientation of samples in the high-pressure cell, which could affect the dehydration product detection sensitivity.
The proximity of our diffusion boundary and the experimental dehydration boundary, particularly their similar pressure gradients, suggests that continuous proton diffusion controls the dehydration process as in antigorite \cite{sawaiDehydrationKineticsAntigorite2013} or diopside with H defects \cite{ingrinDiffusionHydrogenDiopside1995}.

The diffusion boundary previously obtained with PBE-BOMD \cite{heSuperionicHydrogenEarth2018} largely overestimates the superionic transition temperature. It is known that supercell size could affect the superionic diffusivity calculations, often leading to overestimation of the transition temperature \cite{ grasselliInvestigatingFinitesizeEffects2022, potashnikovHighprecisionMolecularDynamics2013}. The short simulation time and the PBE's inaccurate description of the H-bond could also contribute to the overestimation of the diffusion boundary.

It is also possible to investigate the change in protons' diffusivity by inspecting the constant-volume heat capacity, $C_V$, through the $P,T$ range shown in Fig.~S6.
The constant-volume heat capacity, $C_V$, can be calculated from MD ensemble averages \cite{klochkoGeneralRelationsObtain2021} as
\begin{equation}
    N C_V = \frac{\mathrm{var}(E)}{k_\mathrm{B} T^2} = \frac{1}{k_\mathrm{B} T^2} \! \left[\big\langle E^2\big\rangle - \big\langle E\big\rangle^2 \right],
\end{equation}
where $k_\mathrm{B}$ denotes the Boltzmann's constant, $N$ denotes the number of atoms, $\langle E^2 \rangle$ and $\langle E \rangle^2$ denotes the run average of energy $E$ and its square $E^2$.
$C_V$ computed using this method does not exhibit a size effect for simulation cell sizes varying from 128--27,648~atoms in both the solid and diffusive states (Fig.~S6).
$C_V$ of an approximately harmonic solid is expected to plateau at the Dulong-Petit limit, $3 N K_\mathrm{B}$. Deviation from this value suggests strongly anharmonic behavior, in the present case, entering the diffusive regime.

\begin{figure*}[htbp]
    \includegraphics[width=.9\textwidth]{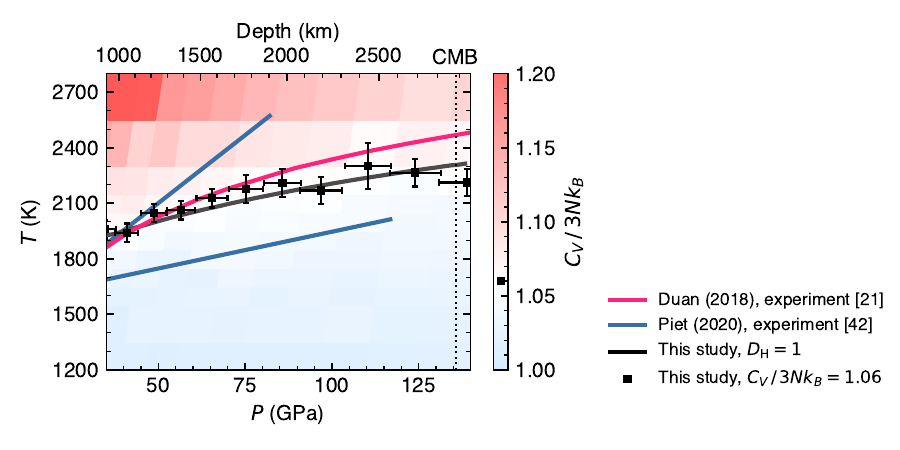}
    \caption{
    The ratio between $C_V$ and the Dulong-Petit limit, $3 N k_B$, at different pressures and temperatures. Black symbols indicate a 6\% excess above the limit. The error bar is determined based on the $P,T$ grid size. Results are compared to the $D_\mathrm{H} = 1$ boundary and measured dehydration phase boundaries \cite{pietDehydrationDAlOOHEarth2020, duanPhaseStabilityThermal2018}.
    }\label{fig:9}
\end{figure*}

Fig.~\ref{fig:9} shows $\delta$'s $C_V$ vs.\ $P,T$ as background color.
For $T < \mbox{1,500}$~K (blueish background), $C_V$ generally follows the Dulong-Petit limit with at most 5\% excess.
For $T >$~1,800~K, $C_V$ increases quickly. Depending on the pressure, the background color turns from blue to white then red between 1,800~K and 2,100~K, indicating  $> 10$\% deviation from $3 N k_B$ within this color change range.
The $D_\mathrm{H} = 1$~\AA$^2$/ns diffusion boundary corresponds to $C_V$ $\sim$6\% above the Dulong-Petit limit.
Measurements of $C_V$ could be useful to determine the onset of $\delta$'s superionic state.

\section{Discussion}
\label{sec:discussion}

This study demonstrates the necessity of adopting a hybrid \textit{ab initio} description aided by deep-learning potential at extreme $P,T$ mantle conditions to address the properties of a hydrous system. The DP-GEN active learning scheme used to develop the potential is highly efficient, requiring only a few thousand massively parallelized DFT calculations on reference configurations, taking only a few days. The number of \textit{ab initio} calculations needed to prepare the interatomic potential for a wide range of $P,T$ conditions, $\sim$3,000 in this case, is far fewer than that of a typical BOMD run necessary for a single $P,T$ sampling, $\sim$$10^4$--$10^5$. This efficiency allows us to use more accurate functionals and perform more accurate simulations using larger simulation cells, longer run times, and denser sampling of points in $P,T$ phase space. Yet, it is still desirable to develop these potentials further to perform under reactive conditions and detect the dehydration process in a simulation.

The current DP-SCAN approach combined with classical MD reproduces measurements of $\delta$'s room temperature compression curve surprisingly well. It paves the way for adopting the path-integral approach to quantum ionic dynamics \cite{tuckermanEfficientGeneralAlgorithms1996, marxInitioPathIntegral1996} in low-temperature simulations of $\delta$. Zero-point motion effects on the EoS of solids at low temperatures, particularly in H$_2$O-ice (e.g., \cite{umemotoNatureVolumeIsotope2015}), are well-known. On $\delta$, the presence of the quantum nuclear effects, e.g., tunneling, is an ongoing debate; evidence supporting its presence \cite{luoInitioInvestigationHbond2022, benoitShapesProtonsHydrogen2005} and absence \cite{trybelAbsenceProtonTunneling2021} has both been published. Such effects could affect H-bond disordering, symmetrization, and details of $P,T$ phase diagrams of hydrous phases (e.g., \cite{lyStabilityDistortionFcc2022}) but have yet to be fully explored in hydrous minerals. At typical mantle temperatures of thousands of Kelvin, classical MD combined with the SCAN functional seems to reproduce closely the thermal EoS of a prototypical lower mantle hydrous phase changing the H-bonds' state.

At mantle and subducting slab conditions (Fig.~\ref{fig:8}), it remains challenging to predict the dehydration process, particularly the transition from the superionic state to the dehydration point. The lack of detailed experimental data in this critical range of conditions aggravates understanding of this process. Electrical conductivity measurements commonly identify diffusion. Electrical conductivity comprises ionic and electronic components: ionic conductivity, according to the Nernst-Einstein relation, will be proportional to free proton concentration, which varies exponentially with temperature; electronic conductivity would require more complex electronic structure calculation, which extends beyond the scope of this study. $\delta$'s electrical conductivity has only been measured up to 1,200~K and up to 20~GPa, and has shown exponential dependence vs.\ reciprocal temperature \cite{wangElectricalConductivityDiaspore2021} similar to ours. Extending such measurements to more extreme conditions is desirable to understand better the relation between the superionic state and the dehydration process.

Simulations of these processes will be fundamental to understanding water circulation in Earth's interior. Subducting slabs carry hydrous phases into the mantle. Describing and predicting the behavior of such phases at extreme conditions is key to understanding mantle dynamics, e.g., volcanism, melt generation, etc. While here we focus on the superionic state, we see a relationship between the onset of the superionic state and the dehydration boundary. Our predicted ``diffusion boundary'' (see Fig.~\ref{fig:8}), the dehydration boundary reported by Ref.~\cite{duanPhaseStabilityThermal2018}, and Ref.~\cite{pietDehydrationDAlOOHEarth2020}'s lower boundary, all share a similar slope that is less steep than the temperature profiles along the subducting slab.
These diffusion and dissociation boundaries intercept the subducting slab geotherm \cite{eberleNumericalStudyInteraction2002, ohtaniHydrousMineralsStorage2015} at a depth of $\sim$2,400--2,700~km near the bottom of the lower mantle, and the normal mantle geotherm \cite{brownThermodynamicParametersEarth1981} at $\sim$1,200--1,500~km, i.e., approximately mid-mantle. These results confirm previous suggestions that $\delta$ could remain stable up to near the bottom of the mantle and only then release water. H$_2$O released from $\delta$ should react with its environment to form other H-bearing phases or melts, e.g., $\epsilon$-FeOOH and FeH$_x$ \cite{terasakiStabilityFeNi2012, pietDehydrationDAlOOHEarth2020} in the deep mantle. Understanding the entire diffusion to dehydration process and the $P,T$ conditions for these transitions will help clarify the deep water cycle, the potential signature of water presence in seismic tomography, and the effects of water on mantle properties.

\section*{Method}
\label{sec:method}

\subsection*{Machine learning potentials}

The NN potential for $\delta$ was developed based on the Deep Potential Smooth Edition (\textsc{DeepPot-SE}) model \cite{zhangEndtoendSymmetryPreserving2018} implemented in \textsc{DeePMD-kit} v2.1 \cite{wangDeePMDkitDeepLearning2018, zengDeePMDkitV2Software2023}. Two-body embedding with coordinates of the neighboring atoms (\texttt{se\_e2\_a}) was used for the descriptor. The embedding network shape is (25, 50, 100). The fitting network shape is (128, 128, 128). The cut-off radius is 6~Å, and the smoothing parameter is 0.5~Å.

The model was trained using the \textsc{Adam} optimizer \cite{kingmaAdamMethodStochastic2017} for $1 \times 10^6$ training steps, with the learning rate exponential decaying from $1\times{10}^{-3}$ to $3.51\times{10}^{-8}$ throughout the training process. The loss function $\mathcal{L}(p_e,p_f)$ is \cite{wangDeePMDkitDeepLearning2018}
\begin{equation}
\mathcal{L}(p_e,p_f) = p_e \lvert\Delta e\rvert^2 + \frac{p_f}{3N} \lvert\Delta f_i\rvert^2,    
\end{equation}
where $p_e$ decays linearly from 1.00 to 0.02, and $p_f$ increases linearly from $1\times10^0$ to $1\times10^3$ throughout the training process.

\subsection*{Active learning scheme}

The DP-GEN concurrent learning scheme \cite{zhangDPGENConcurrentLearning2020} was employed to create the training data set and to generate the potential. We randomly extracted 118 labeled configurations from 59 BOMD runs at various $V,T$s to generate the initial potentials and to kickstart the DP-GEN training process. 6 DP-GEN iterations were performed to explore the configuration space and to eventually generate a potential reaching satisfactory accuracy requirement for DPMD for a temperature range of $300 < T < 3,000$~K. 4~candidate DP potentials initialized with different random seeds were trained in each iteration. They were used to perform $NVT$ DPMD simulations at various $V,T$s for a few thousand timesteps. After the simulations, the error estimator (model deviation), $\epsilon_t$, is calculated every 50 MD steps based on the force disagreement between the candidate DPs \cite{zhangActiveLearningUniformly2019, zhangDPGENConcurrentLearning2020}:
\begin{equation}
\epsilon_t = \max_i \sqrt{\left< \Vert F_{w,i} (\mathcal{R}_t) - \left< F_{w,i} (\mathcal{R}_t) \right> \Vert^2 \right>}\,,
\end{equation}
where $F_{w,i}(\mathcal{R}_t)$ denotes the force on the $i$-th atom predicted by the $w$-th potential for the $\mathcal{R}_t$ configuration. For a particular configuration, if $\epsilon_t$ satisfies $\epsilon_\mathrm{min} \le \epsilon_t \le \epsilon_\mathrm{max}$, the corresponding configuration is collected, then labeled with DFT forces and total energy then added into the training dataset; if $\epsilon_t < \epsilon_\mathrm{min}$, these configurations are considered ``covered'' by the current training dataset; if $\epsilon_t > \epsilon_\mathrm{max}$ are considered failed and were discarded. After a few iterations, almost no new configurations are collected according to this standard ($> 99$\% are ``accurate'' for a few iterations), the DP-GEN process is then complete. The parameters for these DP-GEN iterations are listed in Table~SII in detail.

After these DP-GEN iterations, our training dataset consists of 3,487 configurations labeled with \textit{ab initio} force and energy. DPMD simulations were performed using the LAMMPS code  \cite{thompsonLAMMPSFlexibleSimulation2022} with 0.5~fs timesteps.

\subsection*{DFT calculations}

Training and testing data sets were based on \textit{ab initio} calculations performed with the \textsc{Vienna Ab initio Simulation Package} (VASP)~v6.3 \cite{kresseEfficientIterativeSchemes1996}. The strongly constrained and appropriately normed (SCAN) \cite{sunStronglyConstrainedAppropriately2015} meta-GGA functional with PAW basis sets were adopted. The cutoff energy for the plane-wave-basis set was set to 520~eV. The Brillouin zone sampling for the $2\times2\times4$ supercells (with 16~f.u.\ of $\delta$, or 128~atoms) used a shifted $2\times2\times2$ Monkhorst-Pack $k$-point mesh. BOMD simulations were performed with a timestep of 0.5~fs.

\subsection*{MD simulations}

A dataset for validating the DP potential is created by performing both BOMD and DPMD $NVT$ canonical ensemble simulations on $2\times2\times4$ supercells (128 atoms) with the N\'ose-Hoover (NH) thermostat/barostat \cite{hooverKineticMomentsMethod1996}. These simulations are performed at 6~$T$s ($T =$~300, 600, 1,200, 1,800, 2,400, and 3,000~K) and 4~$V$s (the conventional cell volume $V =$~40, 44, 48, and 54~Å$^3$). This mesh covers the pressure range of $\sim$20--150~GPa. These simulations start from configurations produced after $10^4$ DPMD equilibration timesteps at each given $(V,T)$ and run for a minimum of 1~ns at each $(V,T)$.
The DPMD simulation time scales linearly with the number of atoms (see Fig.~S7), similarly with previous benchmarks \cite{zhangDeepPotentialMolecular2018}.

The investigation of the 300~K compression curve involves DPMD simulations with 1,024-atom (or $4\times4\times8$) supercells. We perform $NPT$ simulations for 50~ps to equilibrate the structure. Then, perform $NVT$ simulations for another 50~ps to obtain the pressure at the given volume.
The investigation of high $P,T$s, involved systematic DPMD simulations with 8,192-atom (or $8\times8\times16$) supercells. We perform $NVT$ simulations. 96 DPMD simulations at different $(P,T)$'s cover the $P,T$ range from 1,500~K to 3,000~K and 10~GPa to 180~GPa (see Fig.~S2). Each MD simulation at equilibrated $V,T$ conditions runs for 2~ns, or $4 \times 10^6$~timesteps. These simulations were performed concurrently on 96~GPUs.

\subsection*{Workflow management}

Our workflows, including the DP-GEN active learning iterations and subsequent MD simulations, were implemented and managed using the Snakemake \cite{kosterSnakemakeScalableBioinformatics2012, molderSustainableDataAnalysis2021} workflow management system.

\section*{Acknowledgments}

DOE Award DE-SC0019759 supported this work. Calculations were performed on the Extreme Science and Engineering Discovery Environment (XSEDE) \cite{townsXSEDEAcceleratingScientific2014} supported by the NSF grant \#1548562 and Advanced Cyberinfrastructure Coordination Ecosystem: Services \& Support (ACCESS) program, which is supported by NSF grants \#2138259, \#2138286, \#2138307, \#2137603, and \#2138296 through allocation TG-DMR180081. Specifically, it used the \textit{Bridges-2} system at the Pittsburgh Supercomputing Center (PSC), the \textit{Anvil} system at Purdue University, the \textit{Expanse} system at San Diego Supercomputing Center (SDSC), and the \textit{Delta} system at National Center for Supercomputing Applications (NCSA).
We gratefully acknowledge Hongjin Wang for the early discussions of the DP method.

\appendix

\bibliography{Geophysics}

\end{document}


\title{\bfseries\textsc{Supplementary Information}\vspace{16pt}\\Probing the state of hydrogen in $\delta$-AlOOH at mantle conditions with machine learning potential}

\author[1]{Chenxing Luo\,\orcidlink{0000-0003-4116-6851}}
\author[1,2]{Yang Sun\,\orcidlink{0000-0002-4344-2920}\thanks{yangsun@iastate.edu}}
\author[1,3,4]{Renata M.\ Wentzcovitch\,\orcidlink{0000-0001-5663-9426}\thanks{rmw2150@columbia.edu}}

\affil[1]{{Department of Applied Physics and Applied Mathematics}, {Columbia University}, {{500 West 120 Street}, {New York}, {10027}, {New York}, {USA}}}
\affil[2]{Department of Physics and Astronomy, Iowa State University, Ames, Iowa 50011, USA}
\affil[3]{{Department of Earth and Environmental Sciences}, {Columbia University}, {{1200 Amsterdam Avenue}, {New York}, {10027}, {New York}, {USA}}}
\affil[4]{{Lamont--Doherty Earth Observatory}, {Columbia University}, {{61 Route 9 West}, {Palisades}, {10964}, {New York}, {USA}}}
\date{}

\maketitle
\thispagestyle{empty}


\textbf{This PDF file includes:}

\begin{itemize}
\item Tables SI and SII
\item Figures S1 to S8
\end{itemize}


\newpage

\normalcolor

\setcounter{section}{0}
\renewcommand{\thesection}{S-\Roman{section}}
\setcounter{figure}{0}
\renewcommand{\thefigure}{S\arabic{figure}}
\setcounter{table}{0}
\renewcommand{\thetable}{S\Roman{table}}

\begin{table}[p]
    \centering
    \renewcommand{\arraystretch}{1.25}
    \caption{Details for RMSE of force and potential energies for validation.}
    \begin{tabular}{c c c c}
    \toprule
        $T$ & $V$ &
        \begin{minipage}{6em}\centering Energy RMSE$/N$ (eV/atom)\end{minipage} & \begin{minipage}{5em}\centering Force RMSE (eV/Å)\end{minipage} \\
        
    \hline
    
        600.0	& 54.0	& 0.001873	& 0.035725 \\
        600.0	& 44.0	& 0.002231	& 0.030642 \\
        600.0	& 40.0	& 0.002494	& 0.031809 \\
        600.0	& 48.0	& 0.002179	& 0.032381 \\
        1200.0	& 44.0	& 0.002140	& 0.047643 \\
        1200.0	& 48.0	& 0.002202	& 0.049318 \\
        1200.0	& 54.0	& 0.001965	& 0.052093 \\
        1200.0	& 40.0	& 0.002487	& 0.052524 \\
        1800.0	& 48.0	& 0.002060	& 0.067983 \\
        1800.0	& 44.0	& 0.001998	& 0.064704 \\
        1800.0	& 54.0	& 0.002045	& 0.070243 \\
        1800.0	& 40.0	& 0.002315	& 0.073160 \\
        2400.0	& 48.0	& 0.002096	& 0.084835 \\
        2400.0	& 44.0	& 0.002090	& 0.084830 \\
        2400.0	& 40.0	& 0.002492	& 0.091925 \\
        2400.0	& 54.0	& 0.001497	& 0.101425 \\
        3000.0	& 48.0	& 0.001981	& 0.122193 \\
        3000.0	& 54.0	& 0.001948	& 0.124198 \\
        3000.0	& 44.0	& 0.001831	& 0.106954 \\
        3000.0	& 40.0	& 0.002675	& 0.119190 \\

    \toprule
    \end{tabular}
    \label{tab:s2}
\end{table}
\clearpage

\begin{table}[p]
    \centering
    \caption{Details of the DP-GEN iterations for generating the training dataset.}
    \renewcommand{\arraystretch}{1.25}
    \setlength{\fboxsep}{5pt}
    \setlength{\fboxrule}{0pt}
    \begin{tabular}{c c c c}
    \toprule
        Iteration & Temperature &
        \fbox{\begin{minipage}{6em}\centering Criteria for selection (eV/Å) \end{minipage}} & 
        \fbox{\begin{minipage}{6em}\centering \# of configurations collected \end{minipage}} \\
    \hline
         Initial & 300 & N/A & 118 \\
         1 & \fbox{\begin{minipage}{6em}\centering 300, 600, 900 \end{minipage}} & $[0.20,0.30]$ & 878 \\
         2 & \fbox{\begin{minipage}{6em}\centering 300, 600, 900 \end{minipage}} & $[0.20,0.30]$ & 547 \\
         3 & \fbox{\begin{minipage}{6em}\centering 300, 600, 900, 1200, 1500 \end{minipage}} & $[0.25,0.35]$ & 554 \\
         4 & \fbox{\begin{minipage}{6em}\centering 300, 600, 900, 1200, 1500, 1800, 2100 \end{minipage}} & $[0.25,0.35]$ & 390 \\
         5 & \fbox{\begin{minipage}{6em}\centering 300, 600, 900, 1200, 1500, 1800, 2100, 2700 \end{minipage}} & $[0.25,0.35]$ & 1000 \\
    \hline
        Total &&& 3487 \\
    \toprule
    \end{tabular}
    \label{tab:s1}
\end{table}
\clearpage

\begin{figure}[p]
    \centering
    \includegraphics[width=\textwidth]{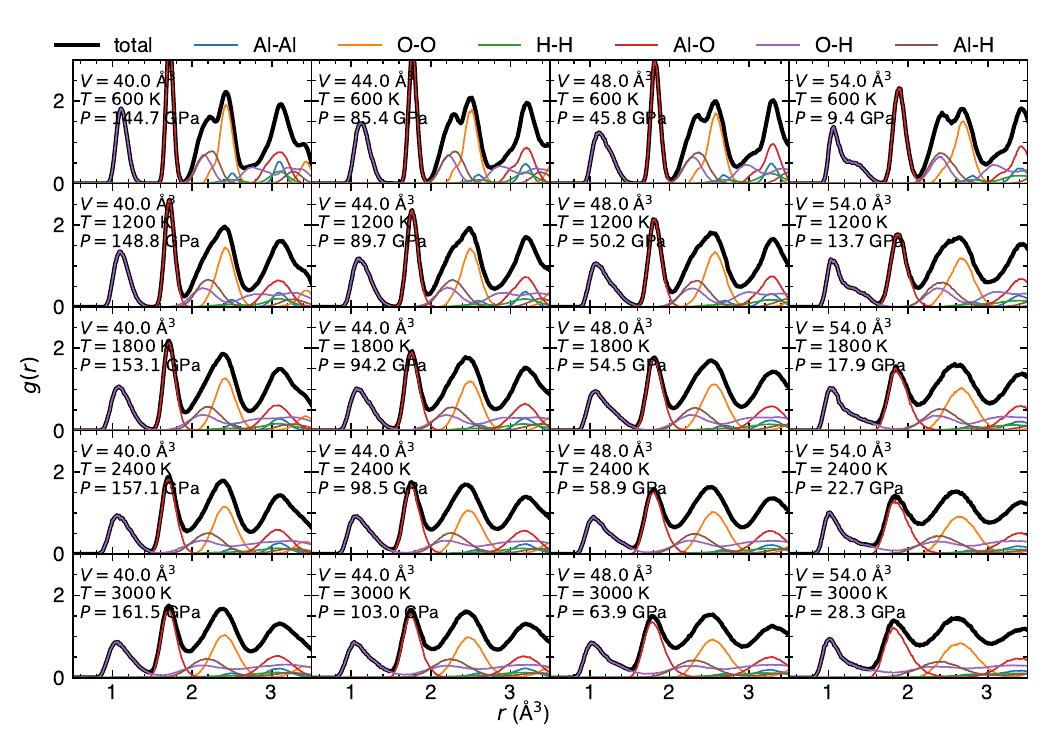}
    \caption{$g(r)$ and partial $g(r)$ at various $V,T$.}
    \label{fig:s1}
\end{figure}
\clearpage

\begin{figure}[p]
    \centering
    \includegraphics[width=.8\textwidth]{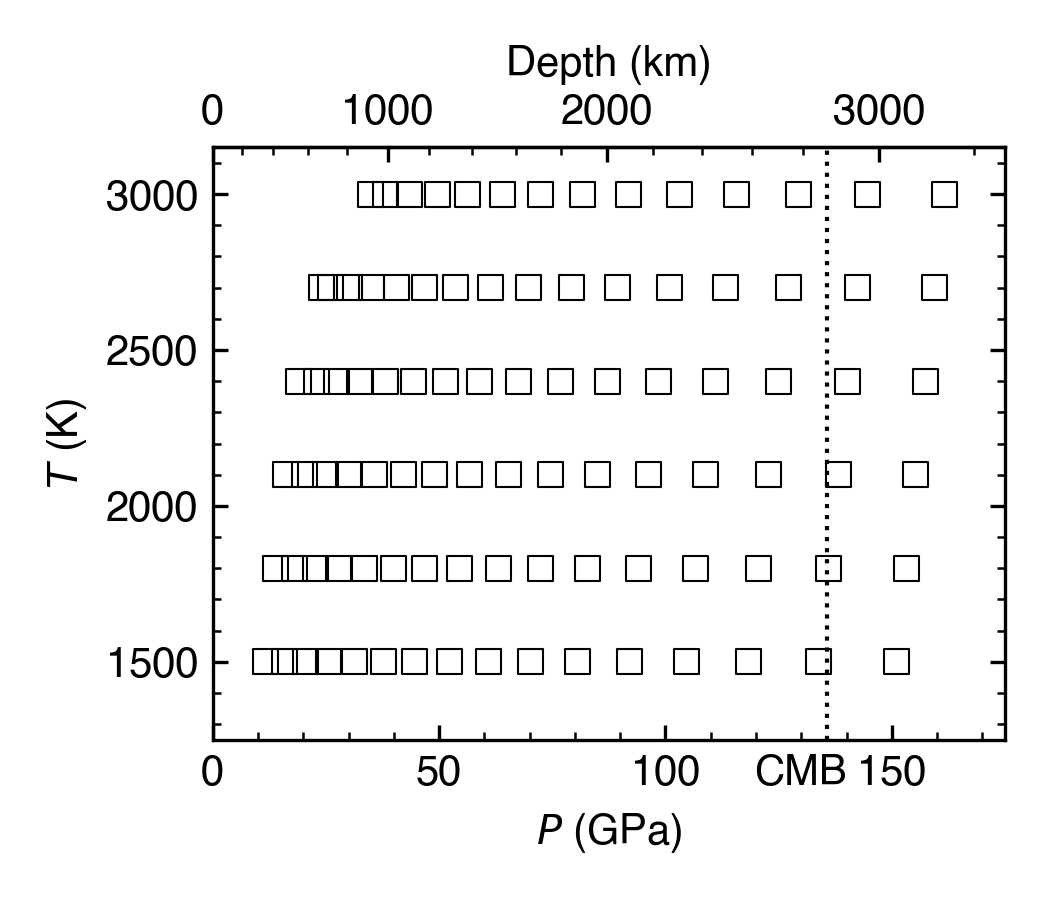}
    \caption{The $(P,T)$'s for systematic DPMD simulations.}
    \label{fig:s2}
\end{figure}
\clearpage


\begin{figure}[p]
    \centering
    \includegraphics[width=\textwidth]{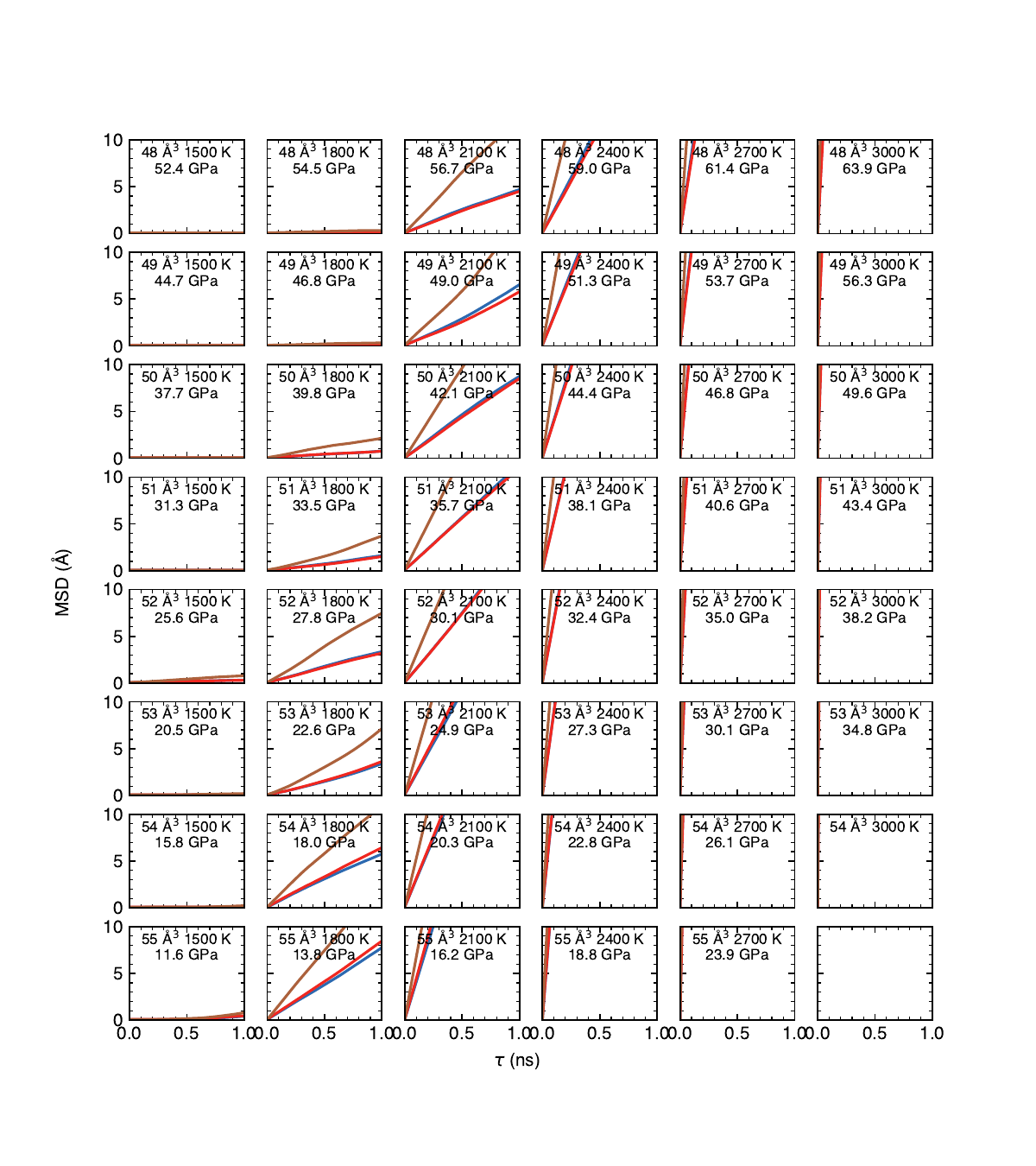}
    \caption{(a) The proton diffusivity at various $V,T$s at $V < 48$~\AA$^3$}
    \label{fig:s3a}
\end{figure}
\clearpage

\begin{figure}[p]
    \ContinuedFloat
    \centering
    \includegraphics[width=\textwidth]{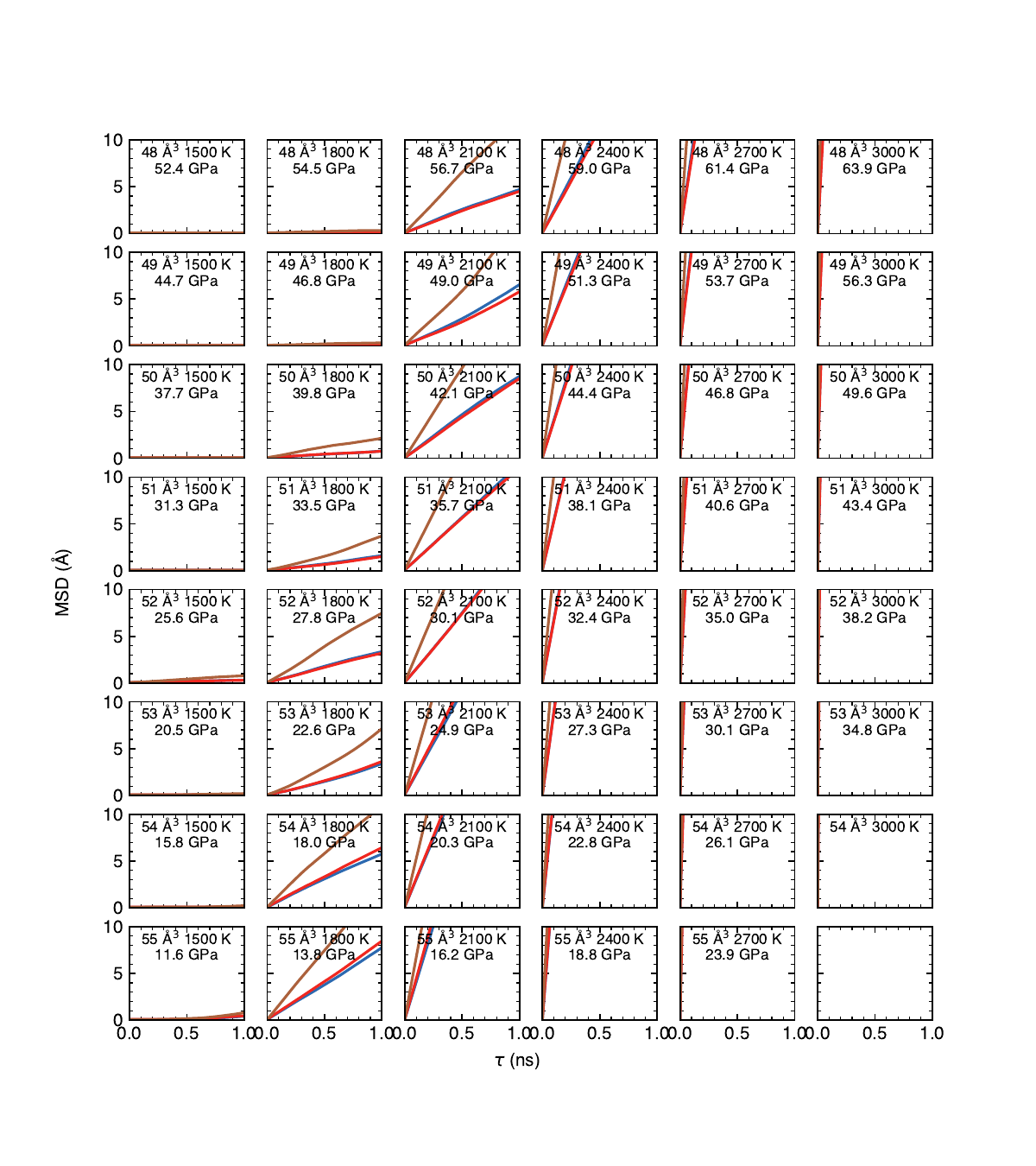}
    \caption{(b) The proton diffusivity at various $V,T$s at $V \ge 48$~\AA$^3$}
    \label{fig:s3b}
\end{figure}
\clearpage

\begin{figure}[p]
    \centering
    \includegraphics[width=.8\textwidth]{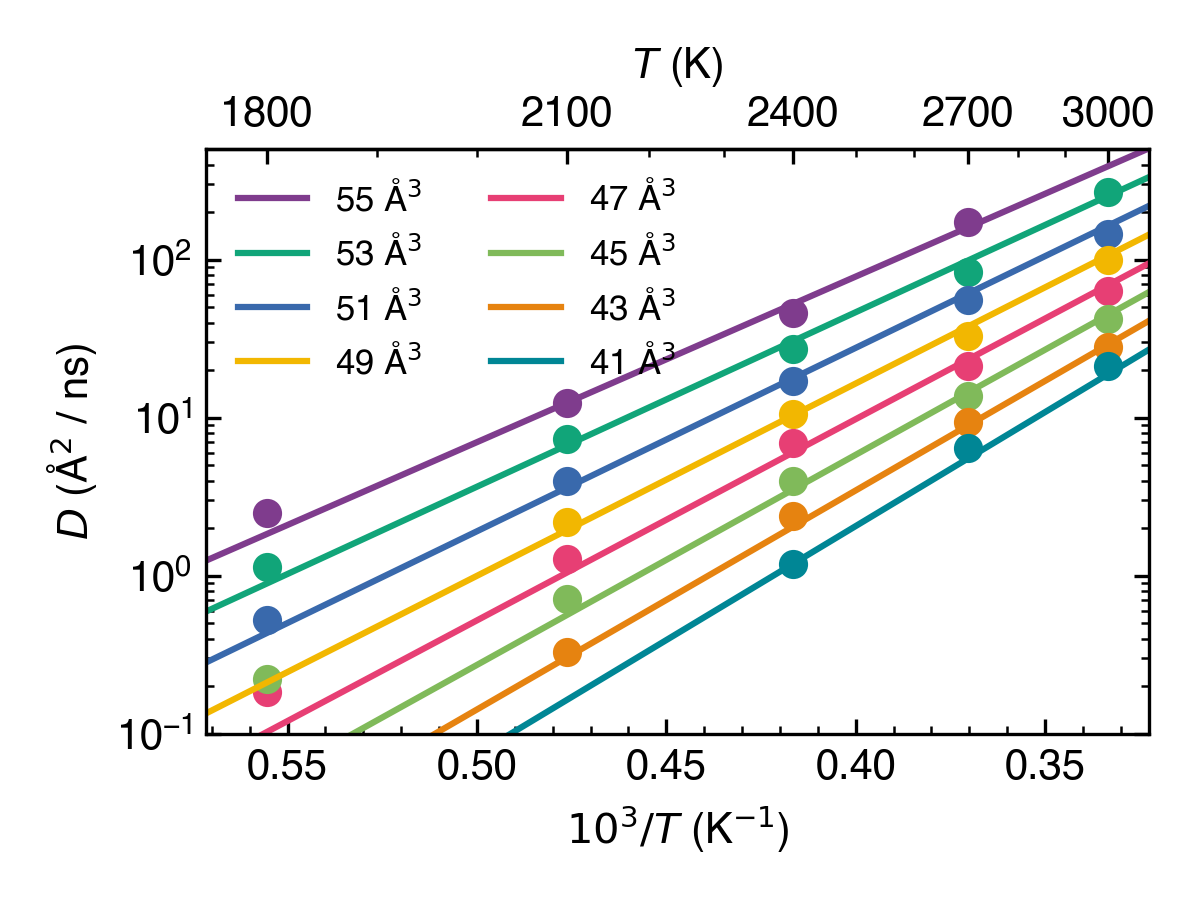}
    \caption{$D_\mathrm{H}$ vs.\ $T$. Solid lines represent the interpolation $\log D_\mathrm{H} (V, T) = a V / T + b / T + c$ at different $V$s.}
    \label{fig:s4}
\end{figure}
\clearpage

\begin{figure}[p]
    \centering
    \includegraphics[width=.75\textwidth]{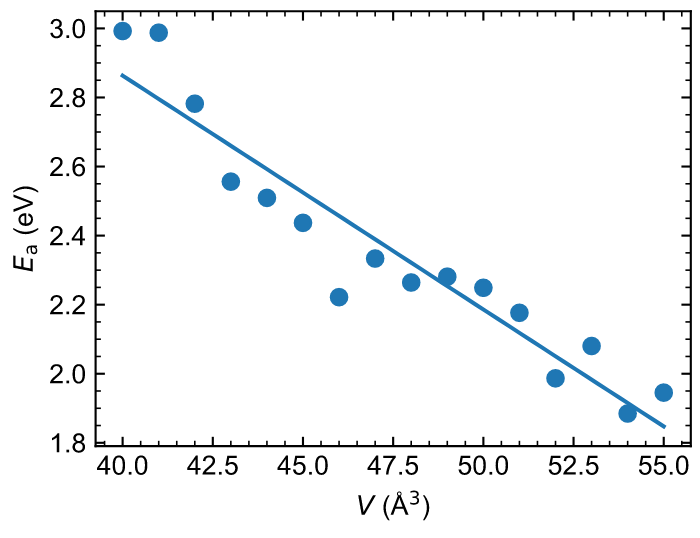}
    \caption{The diffusion activation energy $E_\mathrm{a}$ vs.\ unit-cell volume $V$.}
    \label{fig:s5}
\end{figure}
\clearpage

\begin{figure}[p]
    \centering
    \includegraphics[width=.8\textwidth]{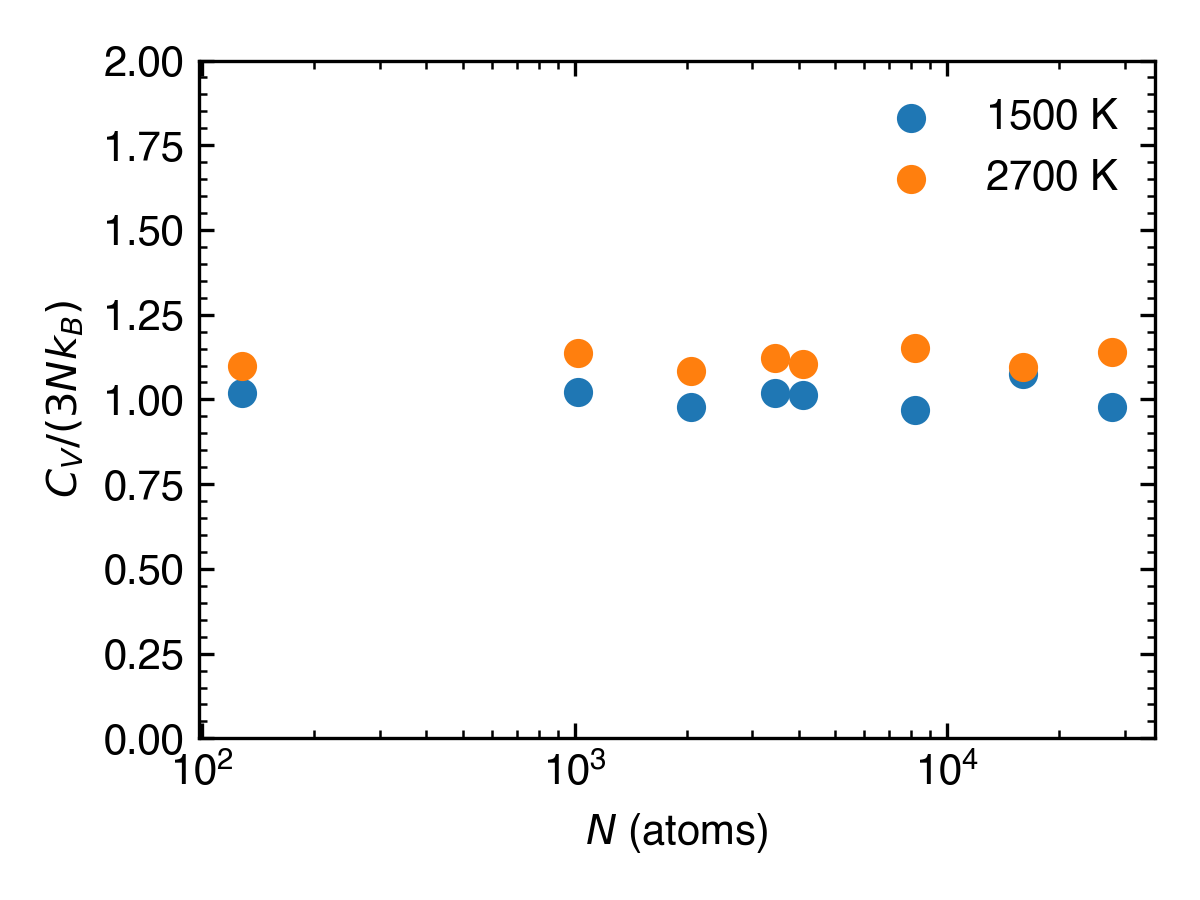}
    \caption{Size effect on $C_V$ at $\sim$110~GPa.}
    \label{fig:s7}
\end{figure}
\clearpage

\begin{figure}[p]
    \centering
    \includegraphics{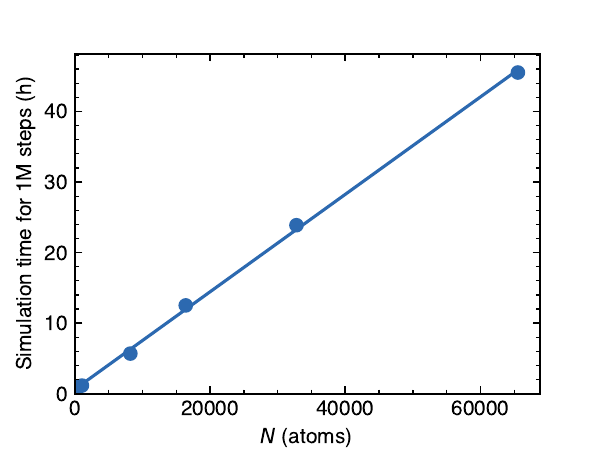}
    \caption{Machine hours for DPMD simulation with $10^6$ MD timesteps vs.\ simulation cell size (128, 256, 512, 1024, 8192, 16,384, 32,768, and 65,536~atoms) on an NVIDIA A100-SXM4-40GB GPU.}
    \label{fig:s7}
\end{figure}
\clearpage
 